\def\papertitle{Vocal Timbre Effects with Differentiable Digital Signal Processing}
\def\paperauthorA{David Südholt}
\def\paperauthorB{Cumhur Erkut}

\documentclass{article}
\usepackage{etoolbox}
\usepackage{arxiv}
\usepackage{amsmath,amssymb,amsfonts,amsthm}
\usepackage{euscript}
\usepackage[T1]{fontenc}
\usepackage[utf8]{inputenc}
\usepackage{ifpdf}
\usepackage[english]{babel}
\usepackage{caption}
\usepackage{subcaption} 
\usepackage{color}

\input glyphtounicode
\newcounter{numauth}
\newcounter{listcnt}
\newcommand\authcnt[1]{\ifdefined#1 \stepcounter{numauth} \fi}

\newcommand\addauth[1]{
\ifdefined#1 
\stepcounter{listcnt}
\ifnum \value{listcnt}<\value{numauth}
\appto\authorslist{, #1}
\else
\appto\authorslist{~and~#1}
\fi
\fi}
\authcnt{\paperauthorB}
\authcnt{\paperauthorC}
\authcnt{\paperauthorD}
\authcnt{\paperauthorE}
\authcnt{\paperauthorF}
\authcnt{\paperauthorG}
\authcnt{\paperauthorH}
\authcnt{\paperauthorI}
\authcnt{\paperauthorJ}
\def\authorslist{\paperauthorA}
\addauth{\paperauthorB}
\addauth{\paperauthorC}
\addauth{\paperauthorD}
\addauth{\paperauthorE}
\addauth{\paperauthorF}
\addauth{\paperauthorG}
\addauth{\paperauthorH}
\addauth{\paperauthorI}
\addauth{\paperauthorJ}

\usepackage{times}

\newif\ifpdf
\ifx\pdfoutput\relax
\else
   \ifcase\pdfoutput
      \pdffalse
   \else
      \pdftrue
\fi

\ifpdf 
  \usepackage[pdftex,
    pdftitle={\papertitle},
    pdfauthor={\authorslist},
    pdfsubject={Proceedings of the 26th International Conference on Digital Audio Effects (DAFx23)},
    colorlinks=false, 
    bookmarksnumbered, 
    pdfstartview=XYZ 
  ]{hyperref}
  \usepackage[pdftex]{graphicx}
\else 
  \usepackage[dvips]{epsfig,graphicx}
  \usepackage[dvips,
    pdftitle={\papertitle},
    pdfauthor={\authorslist},
    pdfsubject={Proceedings of the 26th International Conference on Digital Audio Effects (DAFx23)},
    colorlinks=false, 
    bookmarksnumbered, 
    pdfstartview=XYZ 
  ]{hyperref}
\fi
\usepackage[hypcap=true]{caption}
\title{\papertitle}

\author{
\paperauthorA \thanks{Work performed as an M.Sc. student in Sound and Music Computing at Aalborg University Copenhagen. Accepted for publication in Proceedings of the 26th International Conference on Digital Audio Effects (DAFx23)}\\
    Centre for Digital Music, Queen Mary University of London, London, UK\\
    {\tt \href{mailto:d.sudholt@qmul.ac.uk}{d.sudholt@qmul.ac.uk}}\\
\paperauthorB \\
    Multisensory Experience Lab, Aalborg University Copenhagen, Copenhagen, Denmark \\ {\tt \href{mailto:cer@create.aau.dk}{cer@create.aau.dk}}
}

\begin{document}
\ifpdf 
  \DeclareGraphicsExtensions{.png,.jpg,.pdf}
\else  
  \DeclareGraphicsExtensions{.eps}
\fi

\graphicspath{{figures}} 


\maketitle

\begin{abstract}
We explore two approaches to creatively altering vocal timbre using Differentiable Digital Signal Processing (DDSP). The first approach is inspired by classic cross-synthesis techniques. A pretrained DDSP decoder predicts a filter for a noise source and a harmonic distribution, based on pitch and loudness information extracted from the vocal input. Before synthesis, the harmonic distribution is modified by interpolating between the predicted distribution and the harmonics of the input. We provide a real-time implementation of this approach in the form of a Neutone model. 

In the second approach, autoencoder models are trained on datasets consisting of both vocal and instrument training data. To apply the effect, the trained autoencoder attempts to reconstruct the vocal input. We find that there is a desirable ``sweet spot’’ during training, where the model has learned to reconstruct the phonetic content of the input vocals, but is still affected by the timbre of the instrument mixed into the training data. After further training, that effect disappears.
 
A perceptual evaluation compares the two approaches. We find that the autoencoder in the second approach is able to reconstruct intelligible lyrical content without any explicit phonetic information provided during training.
\end{abstract}

\section{Introduction}

Neural singing voice synthesis has made great progress over recent years. Many efforts are focused on generating natural-sounding voices. The fame of the classic ``vocoder'' sound however, popularized by artists like Daft Punk or Kraftwerk shows the desire for creative timbre manipulation of vocals, where naturalness is not a desired characteristic.

Differentiable Digital Signal Processing (DDSP) \cite{engelDDSPDifferentiableDigital2020} proposes an end-to-end learning approach for neural audio synthesis. Instead of generating signals sample-by-sample in the time domain, or time-varying spectra in the frequency domain, DDSP offers a library of synthesizer components implemented entirely within a framework supporting auto-differentiation. In the case of timbre transfer, an autoencoder model is trained to reconstruct a monophonic sound source based on into pitch and loudness information by generating time-varying control parameters for the synthesizers. The loss is calculated by comparing the spectrogram of the  generated audio from the synthesizers to that of the original audio on multiple timescales. The auto-differentiable implementation allows the gradient of the loss to backpropagate through the synthesizers to update the model weights of the autoencoder.

The synthesizers are based on the spectral modeling synthesis (SMS) \cite{Serra1990} framework. The harmonic components of the sound are generated by a sum of $K$ sinusoids. The decoder predicts $K$ time-varying amplitudes $A_k(n)$, referred to as the \emph{harmonic distribution}, since the sinusoids are defined to oscillate at integer multiples of the (also time-varying) fundamental frequency $f_0(n)$ extracted by the encoder. Thus, the output of the harmonic component $x_h$ can be formulated as

\begin{equation}\label{eq:harmonic}
    x_h(n) = a(n)\sum_{k=1}^K A_k(n)\cdot\sin(\phi_k(n))\,,
\end{equation}
where $a(n)$ is a global amplitude, also predicted by the decoder, and $\phi_k(n) = 2\pi\sum_{m=0}^n kf_0(m)$ is the instantaneous phase of the $k$-th harmonic.

Additionally, the decoder predicts the time-varying magnitude responses of a finite impulse response (FIR) filter. The non-harmonic components of the sounds are generated by processing white noise through these FIRs.

Recent approaches extended the DDSP components and source waveforms. Masuda \cite{NaotakeSynthesizer} proposed a novel approach to synthesizer sound matching by implementing a basic subtractive synthesizer using differentiable DSP modules. Shan \cite{Shan2022Differentiable} introduced Differentiable Wavetable Synthesis (DWTS), a technique for neural audio synthesis that learns a dictionary of one-period waveforms through end-to-end training. Lee \cite{Lee2022Differentiable} formulated recursive differentiable artificial reverberation components, allowing loss gradients to be back-propagated end-to-end, and implemented these models with finite impulse response (FIR) approximations. Finally, Wu \cite{Wu2022DDSP} proposed a new vocoder called SawSing for singing voice, which synthesizes the harmonic part of singing voices by filtering a sawtooth source signal with a linear time-variant finite impulse response filter whose coefficients are estimated from the input mel-spectrogram by a neural network.

Despite these achievements, the use of the classical DDSP for a vocal input with intelligible lyrics has not been explored or exploited, except in  \cite{alonsoExplorationsSingingVoice2021}. In this paper, we propose two methods of adapting DDSP to create vocal effects. We provide a real-time implementation and report perceptual experiments to evaluate our approaches. The structure of this short paper follows our approaches and experiments.

\section{Vocal Effects With DDSP}
Our first approach focuses on altering the predicted synthesizer parameters in a vocoding-inspired manner and will be referred to as the \emph{vocoding approach}. The other makes use of a latent encoding of timbre information and will be referred to as the \emph{latent approach}.

\subsection{Vocoding Approach}

The vocoding approach uses a model trained for timbre transfer. A decoder solely conditioned on pitch and loudness features is trained on audio recordings of a specific instrument, e.g.\ a trumpet. After training has completed, extracting pitch and loudness from any input audio can be used to generate the same melody line in the sound of a trumpet by using the trained decoder to predict corresponding synthesizer controls.

To create the effect of a ``talking trumpet'' from vocal input, we extract pitch and loudness information from the input. Before the synthesis step however, the harmonic distribution $A_k$ is replaced by an altered distribution $A^{\text{out}}_k$ by interpolating between the predicted distribution and the harmonics $A^{\text{in}}_k$ of the input.

To generate $A^{\text{out}}_k$, a user-supplied interpolation factor $p\in[0, 1]$ is introduced. The modified distribution can then be calculated as

\begin{equation}\label{eq:modharmdist}
    A^{\text{out}}_k =
    \begin{cases}
    (1 - p)A_k + A^{\text{in}}_k & kf_0 < \frac{f_s}{2}\\
    0 & \text{ else}
    \end{cases}\,,
\end{equation}
taking care not to include oscillators at frequencies exceeding the Nyquist limit of $f_s/2$. Note that for $p = 0$, $A^{\text{out}}_k = A_k$. In this way, we can create a hybrid harmonic distribution containing both aspects of the timbre of the instrument the decoder was trained on, and the spectral contour of the phonetic content of the vocal input.

A real-time implementation of this approach is made available as a Neutone model at \url{https://github.com/dsuedholt/ddsp_xsynth}.

\subsection{Latent Approach}

In the latent approach, the encoder generates a vector $z$ in addition to pitch and loudness information. We use an encoder provided in the DDSP library that calculates mel-frequency cepstral coefficients (MFCCs) of the input audio at every time step, and processes them through a recurrent layer before projecting them to the latent space.

In this approach, no modifications are applied to the decoder output. Instead, the effect is generated through selection of the training datasets. As explored previously \cite{alonsoExplorationsSingingVoice2021} and confirmed through preliminary experiments, simply training a VAE on recordings of a singing voice can be sufficient to obtain a model capable of reconstructing a vocal input from a different singing voice in the style of the training data with intelligible lyrics.

The idea behind this approach is to add in other monophonic instruments, such as a trumpet or a synthesizer, to the training data, to influence the timbre of the reconstructed vocals in musically interesting ways.

During the experiments, it became clear that if the model is trained until the training loss converges, a decoder with a sufficient number of parameters learns to distinguish between vocal input and the additional instrument, and is able to reconstruct both accurately. This results in a model that is effectively just performing voice transfer. 

\begin{figure}
    \centering
    \includegraphics[width=0.75\columnwidth]{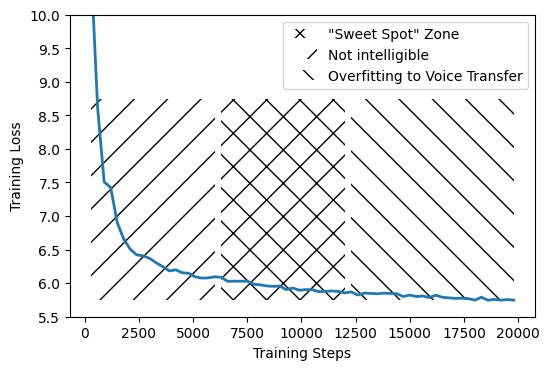}
    \caption{\emph{The training process of a latent encoding model on a combined dataset of vocal performances and brass instruments. Early during training, it cannot reconstruct intelligible lyrics yet. Then it transitions into the ``sweet spot'' where lyrical content is preserved, but the timbre is affected by the additional instrument. After further training, that effect disappears, and the model performs regular voice transfer.}}
    \label{fig:sweetspot}
\end{figure}

However, there appears a ``sweet spot'' early on in training, where the model is already able to reproduce the lyrical content of the input, but has not yet learned to fully distinguish between the different input sources. At this point, the timbre of the reconstructed vocals is affected noticeably by the additional instrument. This is illustrated in Figure~\ref{fig:sweetspot}.

\section{Experiments}

A dataset of vocal performances was created from the Children's Song Dataset (CSD) \cite{choiChildrenSongDataset} and the MUSDB18 dataset \cite{rafiizafarMUSDB18CorpusMusic2017}; instrument datasets were created by combining respective instrument recordings taken from the University of Rochester Multi-Modal Music Performance dataset (URMP) \cite{liCreatingMultitrackClassical2019}. Additionally, a synthesizer performance was obtained by processing randomized MIDI at varying velocities and pitches through a software synthesizer.

Sound examples demonstrating the effect of the vocoding approach at various values for $p$, as well as the "sweet spot" effect of the latent approach, are available at \url{https://dsuedholt.github.io/ddsp-vocal-effects}.

We performed a perceptual evaluation to compare the two approaches. We trained and used the following four models:

\begin{description}
\item[VC-Synth:] Timbre transfer model trained on the synth dataset, vocoding approach, $p=0.7$
\item[VC-Brass:] Timbre transfer model trained on the brass dataset, vocoding approach, $p=0.7$
\item[Z-Vocals:] Latent encoding voice transfer model trained on a single singer from the CSD dataset
\item[Z-Mixed:] Latent encoding model trained on a mixed dataset from the MUSDB18 medley vocals (multiple singers) and the synth dataset

\end{description}

\begin{figure*}[tbph]
    \begin{subfigure}[b]{0.5\columnwidth}
        \includegraphics[width=\columnwidth]{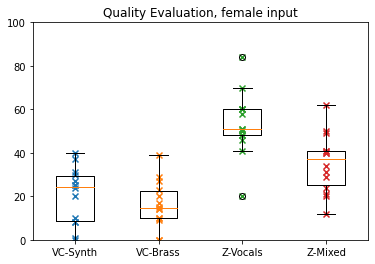}
\end{subfigure}
    \begin{subfigure}[b]{0.5\columnwidth}
        \includegraphics[width=\columnwidth]{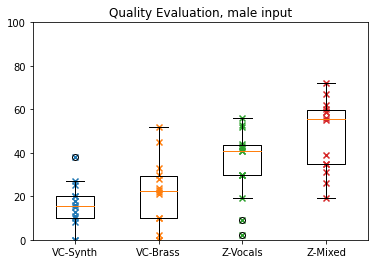}
    \end{subfigure}
    \hfill
    \begin{subfigure}[b]{0.5\columnwidth}
        \includegraphics[width=\columnwidth]{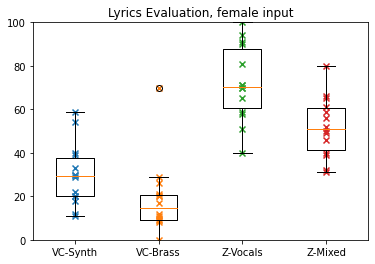}
    \end{subfigure}
    \begin{subfigure}[b]{0.5\columnwidth}
        \includegraphics[width=\columnwidth]{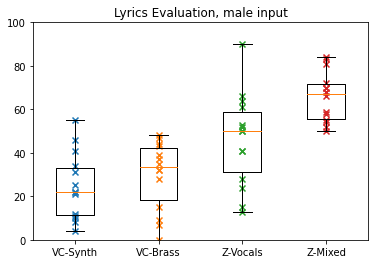}
    \end{subfigure}
    \hfill
    \begin{subfigure}[b]{0.5\columnwidth}
        \includegraphics[width=\columnwidth]{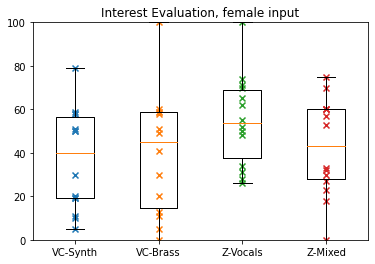}
    \end{subfigure}
    \begin{subfigure}[b]{0.5\columnwidth}
        \includegraphics[width=\columnwidth]{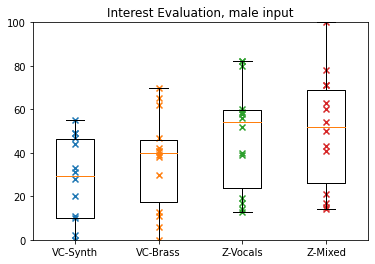}
    \end{subfigure}
    
    \caption{\emph{Results of the perceptual evaluation. All individual ratings are displayed as a scatter plot. A box plot marks the median rating with a horizontal line. The box itself extends from the first to the third quartile of the ratings.}}\label{fig:plots}
\end{figure*}

Two vocal samples, one performed by a male, one by a female singer, were processed by all four models. 15 participants rated the output in a multi-stimulus test under the following three aspects:

\begin{enumerate}
    \item Perceived audio quality
    \item Intelligibility of the lyrics
    \item How musically interesting the effect is
\end{enumerate}
The results of the subjective evaluation are shown in Figure~\ref{fig:plots}.

The clearest result can be found in the rating of the lyrical intelligibility aspect on the female input sample, where the latent encoding models clearly outperform the vocoding models. The same trend, although to a lesser degree, is shown in the evaluation of the male input sample. This seems to confirm that the MFCC + RNN encoder is already capable of reproducing intelligible lyrics without any explicit phonetic information.

None of the models are rated particularly favorably under the aspect of perceived audio quality, although the latent encoding models perform slightly better than the vocoding models. This could potentially be improved by working sampling rates greater than 16 kHz.

The highly subjective rating according to ``musical interest'' shows the highest variance of the ratings, although a slight trend favoring the latent encoding models seems to exist.

\section{Conclusion}
We presented two methods of creating vocal effects that show how the model training and the synthesis stage of the DDSP pipeline can be manipulated for creative effect. We demonstrated that no conditioning on explicit phonetic information is needed to preserve lyrical intelligibility while altering the timbre of the vocal input. These results pave the way towards synthetic "talking" instruments, as well as better understanding of the DDSP training mechanisms and strategies. Still, implementing a unified voice synthesis framework such as NANSY++ \cite{choi2023nansy} remains a future challenge for our field in general.

\bibliographystyle{IEEEbib}
\bibliography{ddspvoxdafx,elicit-results}

\end{document}